\documentclass[fleqn,usenatbib]{mnras}

\usepackage{newtxtext,newtxmath}
\usepackage[T1]{fontenc}

\DeclareRobustCommand{\VAN}[3]{#2}
\let\VANthebibliography\thebibliography
\def\thebibliography{\DeclareRobustCommand{\VAN}[3]{##3}\VANthebibliography}

\usepackage{graphicx}	% Including figure files
\usepackage{amsmath, amsmath}	% Advanced maths commands
\usepackage{soul}
\usepackage{natbib}

\title[Local velocity from cosmic dipole using simulations]{Extracting local velocity from cosmic dipole using simulations}

\author[M. Panwar et al.]{
Mohit Panwar,$^{1}$\thanks{E-mail: mohitpan@iitk.ac.in}
Akash Gandhi,$^{1}$
Pankaj Jain$^{1,2}$
\\
$^{1}$Department of Physics, Indian Institute of Technology, Kanpur 208016, India\\
$^{2}$Department of Space, Planetary \& Astronomical Sciences \& Engineering(SPASE), Indian Institute of Technology, Kanpur 208016, India
}

\date{Accepted XXX. Received YYY; in original form ZZZ}

\pubyear{\the\year{}}

\begin{document}
\label{firstpage}
\pagerange{\pageref{firstpage}--\pageref{lastpage}}
\maketitle

% Abstract of the paper
\begin{abstract}
Our velocity with respect to the cosmic frame of rest leads to a dipole in the number count distribution of galaxies. The dipole depends on the source spectrum, which is usually assumed to be a power law, $S(\nu) \propto \nu^{-\alpha}$ and on the flux dependence of the number density of sources. The latter is also generally assumed to be a power law, parametrised with exponent $x$. The velocity can be extracted from the observed dipole once the two parameters $x$ and $\alpha$ are known. The standard procedure uses the mean value of $\alpha$ across the entire sample, and the parameter $x$ is inferred by fitting the cumulative number count, $\frac{dN}{d\Omega}(>S_*) \propto S_*^{-x}$, near the flux limit $S_*$ of the survey. Here, we introduce a simulation procedure to extract the velocity which directly uses the $\alpha$ values of each source rather than their mean and does not rely on the functional form of the cumulative number count near the flux limit. We apply this to the quasar sample in CatWISE2020 data and find that the final results differ from the standard procedure by approximately one sigma.
\end{abstract}

% Select between one and six entries from the list of approved keywords.
% Don't make up new ones.
% \begin{keywords}
% keyword1 -- keyword2 -- keyword3
% \end{keywords}

%%%%%%%%%%%%%%%%%%%%%%%%%%%%%%%%%%%%%%%%%%%%%%%%%%

%%%%%%%%%%%%%%%%% BODY OF PAPER %%%%%%%%%%%%%%%%%%

\section{Introduction}
The $\Lambda$CDM model is based on the hypothesis that the Universe is statistically homogeneous and isotropic at sufficiently large length scale of order 100 Mpc. This hypothesis is generally known as the Cosmological Principle (CP). The Universe is expected to appear isotropic in a special frame, called the Cosmic Rest Frame (CRF). The relative velocity of the solar system with respect to this frame leads to a dipole anisotropy in the matter distribution and in the temperature field of the Cosmic Microwave Radiation (CMB). The observed dipole anisotropy in the CMB observations \citep{1993ApJ...419....1K, Bennett_2003, Planck_mission, Planck_Dipole} leads to a velocity $(v_{\mathrm{CMB}})$ of $369.82\pm 0.11$km s$^{-1}$ in the direction $(l_{\mathrm{CMB}},\: b_{\mathrm{CMB}})=(264.021^\circ\pm 0.011^\circ,\: 48.253^\circ\pm 0.005^\circ)$\citep{Planck_2018_results}.

The dipole in the matter distribution arises due to the Doppler and aberration effects. In any flux-limited all-sky survey, sources above some limit flux $S_{*}$ are detected. Therefore, sources having flux density in the vicinity of $S_{*}$ will move in or out of the survey depending on the observer's velocity $\mathbf{v}$ and position of the source relative to $\mathbf{v}$. The source spectrum typically follows a power law $S(\nu)\propto \nu^{-\alpha}$, where $\alpha$ is the spectral index and $\nu$ is frequency. The resulting dipole signal in the number count due to these two physical effects is known as the kinematic dipole $\boldsymbol{D}_{\mathrm{kin.}}$ \citep{Baldwin_Ellis_1984}. 
Assuming that the cumulative or integral number count i.e., number of sources per unit solid angle having flux density $S$ greater than the limiting flux density is given by, $\frac{dN}{d\Omega}{\small(S>{S_{*}})}\propto {S_{*}}^{-x}$, then the kinematic dipole is given by,
\begin{equation}\label{eq:kinematic_dipole}
{\mathcal{\boldsymbol{D}}}_{\mathrm{kin.}}=[2 + x(1 + \alpha)]\mathbf{v}/c \,, 
\end{equation}
where $c$ is the speed of light in a vacuum. Here, the factor $2$ arises due to the aberration effect, and the factor $x(1+\alpha)$ due to the Doppler effect.

\begin{figure}
	\includegraphics[width=\columnwidth]{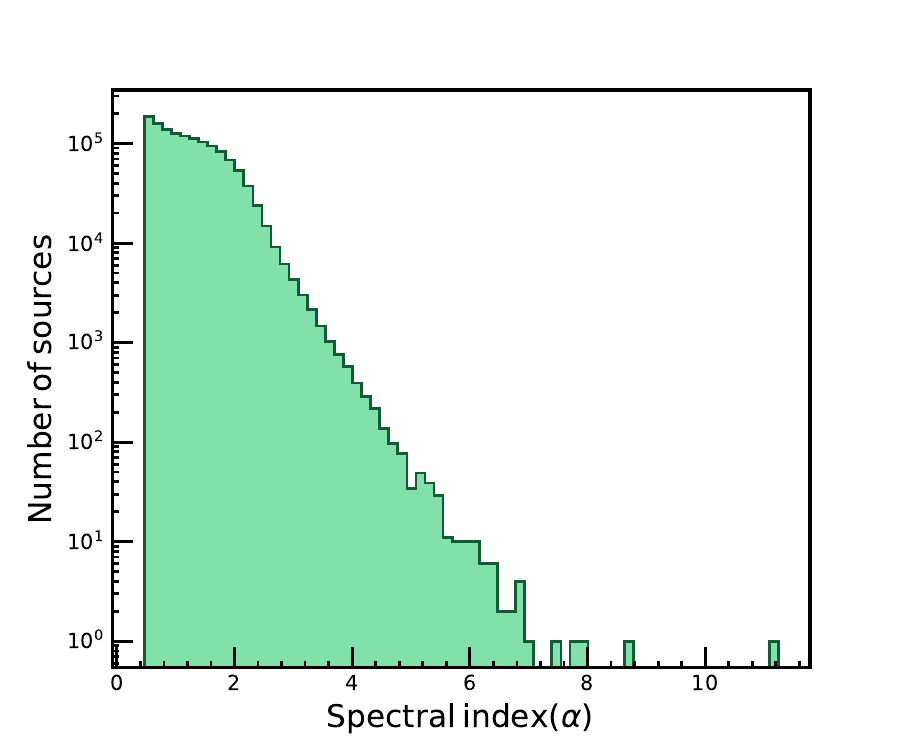}
    \caption{\label{fig:alpha_distribution} The distribution of spectral indices of $1355352$ quasar sample \citep{Secrest_2021}. There are approximately $85\%$ have $\alpha < 2$.}
\end{figure}

The kinematic dipole (equation \ref{eq:kinematic_dipole}) holds if the integral number count near the flux limit and the spectral distribution follow a simple power law. 
It is also assumed that we can simply use the mean value of $\alpha$ in the distribution. Generally, these distributions may not obey a simple power law. Furthermore, in some surveys, we see a wide range of values of $\alpha$. For example, the spectral index $(\alpha)$ distribution comprising the quasar sample in CatWISE2020 data \citep{CatWISE2020_catalogue} is displayed in Fig. \ref{fig:alpha_distribution} \citep{Secrest_2021} and exhibits values from $0.46$ to $11.24$ with an average value of $1.26$.
The distribution has different slopes above and below $\alpha\sim 2$. We point out that it is the sources close to the flux limit that really contribute to the dipole.
Given the large variation in the value of $\alpha$, it is not clear that one can use the sample mean, and the integral number count distribution need not be a simple power law. 

In this paper, we propose a procedure to directly estimate the velocity using simulations rather than equation~\ref{eq:kinematic_dipole}. This eliminates the need to use the mean value of $\alpha$ and to estimate the parameter $x$. The basic idea is that, given our local velocity, we can determine whether a particular source will move in or out of our flux limit. Hence, assuming a value for the local velocity, we can eliminate the contribution of the Doppler effect on our data sample.
The resulting dipole in the data sample will arise entirely due to the aberration effect and must be consistent with the input velocity. This procedure can be iterated till we find a consistent value of the local velocity.

We point out that the NVSS radio sources deviate from a simple power law and instead support a modified differential power law, $\left(\frac{{d}^{2}N}{{dS} \, {d}\Omega}\right) \propto {S}^{-1-x-\beta\ln{S}}$ \citep{TIWARI20151}. Similarly, a power law does not provide a good fit to the full CatWISE2020 quasar population \citep{Panwar_2024}. In principle, one only requires the power law fit near the flux limit.  However, if the full data do not show a power law distribution, the results may depend on the upper cut imposed, and also the error in the extraction of parameter $x$ may increase due to the decrease in the number of sources. We point out that there is another interesting observable, namely flux-weighted number count, which samples the entire distribution rather than the sources close to the flux limit \citep{Singal_2011,TIWARI20151}.

Observationally, the matter dipole has proven to be more complicated. The expected dipole anisotropy in number count, estimated from various flux-limited sky survey catalogues that span the frequency range from radio to infrared, has been found to be in disagreement with kinematic interpretations \citep{10.1046/j.1365-8711.1998.01536.x, Singal_2011, TIWARI20151, Rubart, Bengaly_2018, Secrest_2021}. The dipole direction roughly matches the CMB dipole, but the amplitude is significantly larger than the CMB-inferred dipole amplitude. The highest departure from the $\Lambda$CDM cosmology was reported at infrared frequency at approximately $ 4.9\sigma$ significance level \citep{Secrest_2021}. The results also indicate a redshift dependence of the dipole \citep{Panwar_2024}. These observations indicate potential deviations from the widely accepted $\Lambda$CDM model of the Universe. Furthermore, these are not the only observations that indicate such deviation; several others, including those related to the large-scale structure (LSS) and the Cosmic Microwave Background (CMB) radiation, also suggest discrepancies. Some of these observations include the dipole anisotropy in the radio polarization offset angles \citep{Jain1998}, alignment of the radio polarizations \citep{Tiwari_Jain_2013,Pelgrims_Hutsemekers_2015,Tiwari_Pankaj_2016}, alignment of the radio galaxy axes \citep{Taylor,Panwar}, anisotropy in the Hubble constant \cite{PhysRevD.105.103510}, large-scale bulk flow observations \citep{Kashlinsky_2008,10.1093/mnras/stad1984}, hemispherical power asymmetry in the CMB \citep{Eriksen_2004_hemispherical_power_asymmetry}, alignment of quadrupole$(\ell=2)$ and octopole$(\ell=3)$ harmonics \citep{Quadrupole_and_octopole_alignment,Octo_Quadrupole_align} and dipole modulation in the CMB polarization \citep{Ghosh_2016}. The observed deviations from  $\Lambda$CDM cosmology are reviewed in \citep{Challenges_for_Lambda-CDM_An_update,Kumar}.

The paper is organised as follows: In Section~\ref{sec:Data}, we briefly discuss the CatWISE2020 quasar sample used for the present study. Section~\ref{sec:theory} outlines the spherical harmonic decomposition of number count in a Cartesian basis along with the $\chi^{2}$ statistic to extract the required dipole anisotropy. In section~\ref{sec:Simulation}, we describe our algorithm to remove the Doppler effect from the quasar sample. In section~\ref{sec:results}, we discuss the results and conclude in section~\ref{sec:conclusion}.

\section{Data}\label{sec:Data}
The present paper uses the quasar samples selected from the CatWISE2020 catalogue \citep{CatWISE2020_catalogue} of infrared sources in $\mathrm{W}1$ and $\mathrm{W}2$ band centred at $3.4\mu$m and $4.6\mu$m, employing a mid-colour cut criterion $\mathrm{W}1-\mathrm{W}1\ge 0.8$ \citep{Stern_2012,Mateos_2012,Secrest_2015} to filter out the most probable quasar candidates. To make this quasar sample appropriate for analysis, several corrections, masks and cuts are applied. Our quasar sample is identical to the sample used by \cite{Secrest_2021} in terms of the masking and cuts implemented to mitigate the known obvious systematics present in the data. The only exception involves the removal of observed inverse linear number density as a function of absolute ecliptic latitude. This has been relaxed since we find that including a quadrupole in our model function directly accounts for this dependence. This inverse linear trend is believed to be correlated with the sky scanning pattern of the Wide-field Infrared Survey Explorer(WISE) satellite, which scans the sky along a great circle centred at the Sun, from the north ecliptic pole to the south ecliptic pole \citep{Wright_2010}. Despite redundant sky coverage at ecliptic poles, which leads to an increase in sensitivity, the observed inverse linear trend is quite abnormal. A plausible explanation, yet to be confirmed with more careful analysis, has been provided in \cite{Secrest_2022}. We directly extract this inverse linear trend in the form of quadrupole anisotropy, along with the desired dipole anisotropy in the number count. Indeed, the quadrupole axis is found to be aligned with ecliptic poles \citep{Kothari_2024}. After all the cuts, we are left with $1355352$ quasars in our sample.

\section{Theory}\label{sec:theory}
We assume the following functional form of the model of number count per pixel \citep{Kothari_2024} varying over the sky,
\begin{eqnarray}\label{eq:number_counts_model}
    N_{\mathrm{model}}(\theta, \phi)=\mathcal{M}_{o}+n_{x}\mathcal{D}_{x}+n_{y}\mathcal{D}_{y}+n_{z}\mathcal{D}_{z} + \nonumber\\
    \mathcal{Q}_{xy}n_{x}n_{y}+\mathcal{Q}_{zy}n_{y}n_{z}+\mathcal{Q}_{xz}n_{x}n_{z}+\nonumber \\
    \mathcal{Q}_{x^{2}-y^{2}}(n_{x}^{2}-n_{y}^{2}) + \mathcal{Q}_{z^2}(3n_{z}^2-1).
\end{eqnarray}
    where $\mathcal{M}_{o}$ is monopole, $(\mathcal{D}_{x}, \mathcal{D}_{y}, \mathcal{D}_{z})$ are dipole and $(\mathcal{Q}_{xy}, \mathcal{Q}_{zy}, \mathcal{Q}_{xz}, \mathcal{Q}_{x^{2}-y^{2}}, \mathcal{Q}_{z^2})$ quadrupole components respectively.
Since we are working in the Galactic coordinate system therefore $(n_{x},\: n_y,\: n_z) = (\cos\theta\cos\phi,\: \cos\theta\sin\phi,\: \sin\theta)$ is a unit vector.
The model parameters are collectively represented as $\boldsymbol{\theta}=\{\mathcal{M}_{o}, \mathcal{D}_{x}, \mathcal{D}_{y}, \mathcal{D}_{z}, \mathcal{Q}_{xy}, \mathcal{Q}_{zy}, \mathcal{Q}_{xz}, \mathcal{Q}_{x^{2}-y^{2}}, \mathcal{Q}_{z^2} \}$.
We estimate the parameters $(\boldsymbol{\theta})$ employing $\chi^{2}$ minimization which is given by,
\begin{equation}\label{eq:ChiSqr}
    \chi^{2}(\boldsymbol{\theta}) = \sum_{i=1}^{\mathrm{N}_{\mathrm{pix}}} \frac{[N_{\mathrm{obs},i}(\theta,\phi) - N_{\mathrm{model},i}(\theta,\phi)]^2}{N_{\mathrm{obs},i}}\,
\end{equation}
where $N_{\mathrm{obs},i}(\theta,\phi)$ is the observed number count in $i_{th}$ pixel with direction specified explicitly by angular coordinates $(\theta,\phi)$. %and  $N_{\mathrm{model},i}(\theta,\phi)$ is the number counts value from eq. \ref{eq:number_counts_model}.
From these best optimal subset of parameters $\boldsymbol{\theta}_{\mathrm{best}} = \{\mathcal{M}_{o}, \mathcal{D}_{x}, \mathcal{D}_{y}, \mathcal{D}_{z}\}$, one can estimate the dipole amplitude,
\begin{equation}\label{eq:dipole_amplitude}
    \mathcal{D} = \frac{\sqrt{\mathcal{D}_{x}^2+\mathcal{D}_{y}^2+\mathcal{D}_{z}^2}}{\mathcal{M}_{o}}\, 
\end{equation}
and the dipole direction $(l,\:b)$,
\begin{equation}\label{eq:dipole_direction}
    l= \tan^{-1}\frac{\mathcal{D}_y}{\mathcal{D}_x},\:\:\: b=\sin^{-1}\frac{\mathcal{D}_z}{\mathcal{D}}.
\end{equation}

To generate the observed number count map, we distribute the quasar candidates into $49152$ equal-area pixels corresponding to $\texttt{nside}=64$, employing the $\texttt{HEALPix}$ pixelization layout \citep{Healpix, healpy_package}. The sky model parameters $(\boldsymbol{\theta})$ have been extracted by performing $\chi^{2}(\boldsymbol{\theta})$ minimization. The one-sigma uncertainty associated with the model parameters $\boldsymbol{\theta}$ are estimated from the covariance matrix, which is given by the inverse of the matrix $B_{kl}=\frac{1}{2}\frac{\partial^{2}\chi^{2}(\boldsymbol{\theta})}{\partial\theta_{k}\partial\theta_{l}}$. From the best optimal parameters, we estimate the dipole amplitude $\mathcal{D}$ and direction $(l,\:b)$ using equation~\ref{eq:dipole_amplitude} and \ref{eq:dipole_direction} and associated uncertainties from the error propagation.

The number count dipole and quadrupole anisotropy results are extensively discussed in \cite{Kothari_2024}. The quadrupole anisotropy is found to be correlated with the Ecliptic poles and is attributed to the WISE observation strategy \citep{Wright_2010}. In this paper, we are only interested in the dipole anisotropy; thus, eliminating the quadrupole anisotropy does not impact the results significantly. Therefore, we removed the best-fit quadrupole anisotropy from the data.

\section{Simulation}\label{sec:Simulation}
This section outlines the simulation steps to extract the velocity. In this case, the number count model can be expressed as,
\begin{equation}
    N_{\rm model}(\theta, \phi)=\mathcal{M}_{o} + n_{x}\mathcal{D}_{x}+n_{y}\mathcal{D}_{y}+n_{z}\mathcal{D}_{z}\,
\end{equation}
where $\boldsymbol{\theta}=\{\mathcal{M}_{o}, \mathcal{D}_{x}, \mathcal{D}_{y}, \mathcal{D}_{z}\}$ are the model parameters. We point out that the quadrupole has been removed from the data.

The simulation proceeds by the following steps:
\begin{itemize} 
\item[(i)] Select a velocity $\mathbf{v}$ directed towards the direction $(l_{\mathrm{input}},\: b_{\mathrm{input}})$.
\item[(ii)] Calculate the rest flux $S_{\mathrm{rest}}$ of each source using the equation $S_{\mathrm{obs}} = S_{\mathrm{rest}}\big[1 + (\mathrm{v}/c)\cos(\psi)\big]^{1+\alpha}$, where $\psi$ is angle between the dipole vector and source's position in the sky. 
\item[(iii)] Select the sources with rest flux density greater than the rest-frame flux density cut $S_{\rm rest}^{\prime}$ which is greater than the lower flux cut $S_{\rm cut}$ present in the quasar sample. A reasonable lower flux cut $S_{\rm cut}$ in the quasar sample is $0.085$ mJy which we use in our analysis. We point out that the Doppler effect has now been taken out of the data and now its dipole anisotropy would arise solely due to the aberration effect.
%We consider the rest frame flux cut of $0.085$ mJy to ensure the sources remain in the catalogue so that the aberration effect solely dominates the dipole anisotropy.   
\item[(iv)] Obtain the dipole parameters $\boldsymbol{\theta}$ from the resulting data. 
\item[(v)] For the selected velocity $\mathbf{v}$ and direction $(l_{\mathrm{input}},\: b_{\mathrm{input}})$, if dipole amplitude ${\rm \mathcal{D}}_{\rm data}=2\mathrm{v}/c$ and $l_{\mathrm{input}}=l_{\mathrm{data}},\: b_{\mathrm{input}}=b_{\mathrm{data}}$ then stop and note the parameters $\boldsymbol{\theta}$. Otherwise, repeat from the step $(\rm i)$ with new velocity.
\end{itemize}
\begin{figure*}
    \includegraphics[width=\textwidth]{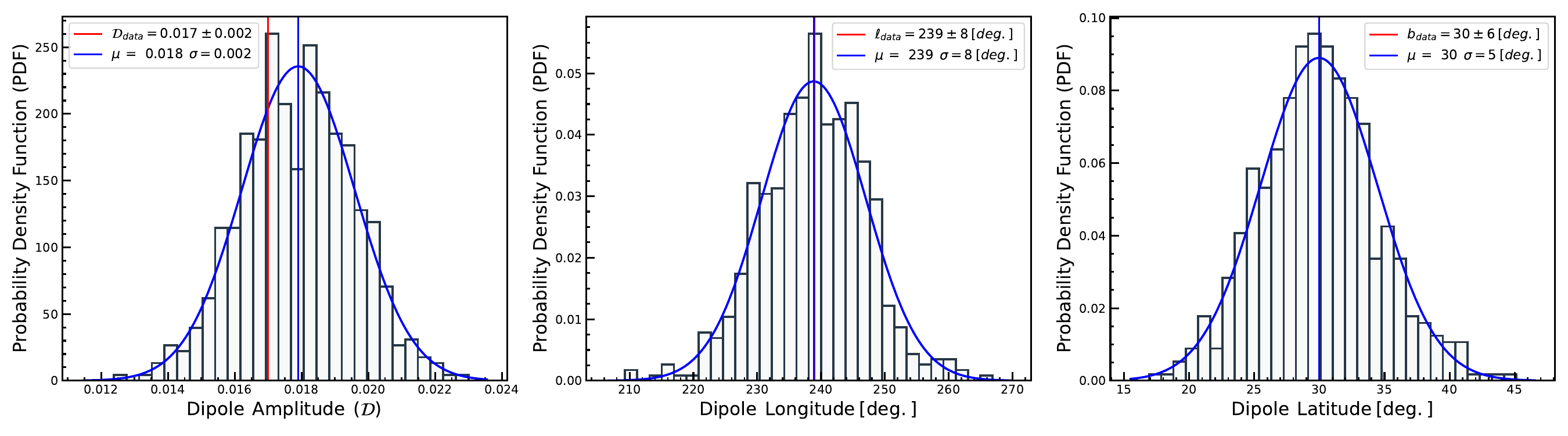}
    \caption{\label{fig:mock_catalogues_results} The plot represents the dipole amplitude $\mathcal{D}$ and direction $(l,\:b)$ distributions of 600 mock catalogues generated with a velocity of 884 km s$^{-1}$ directed towards $(l,b) = (239^\circ,\:31^\circ)$. This figure shows that the velocity and direction at the nominal rest-frame flux density cut $S^{\prime}_{\rm rest} = 0.088$ mJy, obtained using the simulation procedure, properly capture the observed dipole (shown in red) in the full quasar sample. The slight difference between the observed dipole amplitude $\mathcal{D}_{\rm data}$ and the mean dipole amplitude from the mock catalogues, inferred by fitting a Gaussian distribution, may primarily be due to the impact of shot noise and the applied mask. Nevertheless, the two dipole amplitudes match well within the error bars.} 
\end{figure*}
\begin{table}
    \caption{\label{tab:SecondSimulationResults}The velocity $v$ and direction $(l,\:b)$ are extracted employing the simulation procedure (Section~\ref{sec:Simulation}) at various rest-frame flux density cuts i.e. sources with rest flux density $>S^{\prime}_{\mathrm{rest}}$[mJy] are considered.}
    \begin{tabular}{lccccr}
        \hline
        S.No.&$S^{\prime}_{\mathrm{rest}}[\mathrm{mJy}]$ &$l$&$b$ & $\mathbf{v}[\mathrm{km\:s^{-1}}]$ &$\chi^{2}/\mathrm{d.o.f}$\\
        \hline
        $1$&$0.0880$ & $238^\circ\pm 8^\circ$& $31^\circ\pm 6^\circ$ & $884\pm 105$&$1.247$\\
        $2$&$0.0888$ & $235^\circ\pm 8^\circ$& $30^\circ\pm 6^\circ$ & $890\pm 105$&$1.231$\\
        $3$&$0.0896$ & $238^\circ\pm 8^\circ$& $30^\circ\pm 6^\circ$ & $881\pm 104$&$1.228$\\
        $4$&$0.094$  & $235^\circ\pm 8^\circ$& $34^\circ\pm 7^\circ$ & $820\pm 97$&$1.163$\\
        $5$&$0.100$  & $229^\circ\pm 8^\circ$& $36^\circ\pm 7^\circ$ & $840\pm 100$&$1.009$\\
        \hline
    \end{tabular}
\end{table}

\subsection{Simulating mock catalogues}
To evaluate the performance of the simulation procedure proposed in Section \ref{sec:Simulation}, we first test it on a simulated mock catalogue. We simulate $N_{iso}$ statistically isotropic random vectors on a unit sphere to generate the mock catalogue. Each source is assigned a flux density $S$ and a spectral index $\alpha$, drawn randomly from the observed distributions to ensure an exact representation. We then apply the relativistic aberration effect and flux modulation for each source in the observer's frame, which is moving with velocity $\mathbf{v}$. The resulting mock catalogue is masked using the same mask as the original catalogue. Finally, we apply the same lower flux density cut as in the original catalogue and retain the same number of simulated sources as the original contains. 

\section{Results and Discussion}\label{sec:results}
In \cite{Kothari_2024}, the authors obtained the dipole amplitude $\mathcal{D}_{\mathrm{kin.}} =(1.7\pm 0.2)\times 10^{-2}$ and direction $(l,\:b)=(239^\circ \pm 8^\circ,\: 30^\circ\pm 6^\circ)$ in Galactic coordinate system at the flux density cut $S_{\rm cut}=0.085$ mJy in the CatWISE2020 quasar sample. To estimate the velocity from the observed dipole using equation \ref{eq:kinematic_dipole}, we need the parameters $\alpha$ and $x$. The exponent $x$ is estimated by considering the quasars for which the integral number count follows a power law near the flux limit. We find that sources with flux density $S < 0.1$ mJy provide a better fit. The exponent $x$ is estimated by maximizing the log-likelihood function which is given by $\mathcal{L}({\rm data}|x)=-N\log\big\{{(S_{\rm min}^{-x}-S_{\rm max}^{-x})/x}\big\}-(x+1)\sum_{i=1}^{N}\log S_{i}$, where ${\rm data}$ stands for the flux densities of the sources, $N$ is number of sources, and $S_{\rm min}$ and $S_{\rm max}$ are the minimum and maximum flux densities in the sample respectively \citep{Ghosh_2017, Panwar_2024}. We obtain the value of $x=2.26\pm 0.04$. Given the likelihood function, $L({\rm data}|x)=\exp\{\mathcal{L}({\rm data}|x)\}$, we derive the posterior probability density function, $P(x|{\rm data})$, employing Bayes' theorem, i.e., $P(x|{\rm data})\propto L({\rm data}|x)P(x)$, where $P(x)$ is the prior distribution function, and the proportionality constant is the normalization constant, which can be obtained using the condition $\int P(x|{\rm data})dx = 1$. By maximising the resulting posterior PDF with a flat prior, we obtain a value for $x$ that is very close to the one obtained by maximising the log-likelihood function. Using the values of the mean spectral index ${\alpha}$ and $x$, we determine the velocity of the Solar System to be $ 717 \pm 85 $ km s$^{-1}$, which deviates from the velocity inferred from the CMB dipole with a statistical significance of approximately $ 4.1\sigma $. 

The simulation procedure presented in Section \ref{sec:Simulation} estimates a velocity of $884$ km s$^{-1}$ towards the direction $(l,\:b) = (238^\circ,\:31^\circ)$ at a rest-frame flux cut of $0.088$ mJy. {The main reason for selecting sources with a rest-frame flux density greater than \(S^{\prime}_{\rm rest}=0.088\) mJy is to ensure that their observed flux density remains above the lower flux cut \(S_{\rm cut} = 0.085\) mJy, where the Doppler effect is expected to have a significant contribution, so that the sources remain in the catalogue.} We apply more stringent rest-frame flux cuts and estimate the velocity and dipole direction accordingly. These results are presented in Table \ref{tab:SecondSimulationResults}. It is apparent from the table that the velocities remain nearly the same at all flux density cuts, and the dipole direction consistently points in the same direction. 

To assess the uncertainty associated with the extracted velocity $\mathbf{v}$ and direction $(l,\:b)$ at a given rest-frame flux density cut, we generate mock catalogues by boosting the observer with the corresponding velocity towards the direction $(l,\:b)$. From these mock catalogues, we extract the velocity and direction following our simulation procedure as outlined in Section~\ref{sec:Simulation} to generate the distributions.

We generated $600$ mock random catalogues using a velocity of $884$ km s$^{-1}$ directed towards $(l,\:b) = (238^\circ,\:31^\circ)$. We obtain the dipole amplitude and direction by employing $\chi^{2}$ minimisation (Eq. \ref{eq:ChiSqr}) for each mock catalogue. Figure \ref{fig:mock_catalogues_results} illustrates the distributions of the dipole amplitude and direction parameters obtained from these catalogues. It is evident from the plot that the velocity and direction parameters inferred through the simulation procedure at \( S^{\prime}_{\text{cut}} = 0.088 \) mJy effectively capture the observed dipole signal in the full quasar sample, with a slightly lower percentage error \( \left(\frac{\Delta\mathcal{D}}{\mathcal{D}}\right)_{\rm mock} \times 100 = (0.00169/0.0179) \times 100 \approx 9.44\% \) in the dipole amplitude, compared to the percentage error inferred from the full quasar sample, \( \left(\frac{\Delta\mathcal{D}}{\mathcal{D}}\right)_{\rm data} \times 100 \approx 11.85\% \). This is to be expected since all sources of error present in the quasar sample, such as the uncertainties associated with flux density and the variation of alpha for each source, have not been taken into account while simulating the mock catalogues. Therefore, this leads to an underestimation of the error associated with the extracted velocity. Hence, we scale the percentage error in velocity to account for this small difference in error in the dipole amplitude.

Finally, we compare the result obtained at the nominal rest-frame flux density cut of $S^{\prime}_{\text{cut}} = 0.088$ mJy with those estimated using equation \ref{eq:kinematic_dipole} and find that the velocity is larger by $1.2\sigma$. Additionally, the statistical significance of deviation from the velocity inferred from the CMB dipole is approximately $4.9\sigma$. Hence, we find that the final velocity obtained by the simulation procedure deviates significantly from that obtained directly by use of equation \ref{eq:kinematic_dipole}.

\section{Conclusions}\label{sec:conclusion}

We propose a simulation procedure to extract the velocity of the Solar System from an all-sky survey catalogue. In the standard procedure, the velocity is estimated from the observed dipole using the expression $\mathcal{\boldsymbol{D}}_{\rm kin} = [2 + x(1 + \alpha)]\mathbf{v}/c$, which incorporates the mean value of the spectral index $\alpha$ for the entire sample and assume that the integral number count near the flux limit follow the power law with exponent $x$. In the observed sample, the $\alpha$ values often show a broad distribution, and it may not be justified to use a mean value. Furthermore, the integral number count may not precisely follow a power law. These two requirements may not hold in general for a wide range of continuum all-sky surveys, which catalogue sources with a wide range of spectral indices. Our simulation procedure explicitly uses the spectral indices of individual sources and does not rely on the functional form of the integral number count near the flux limit. We implement the simulation procedure on the CatWISE2020 quasar sample to extract the velocity of the Solar System and find that the velocity is approximately $1.2\sigma$ larger than the value inferred using the equation~\ref{eq:kinematic_dipole}. The error in the extracted velocity can be determined by simulations. We find that our results for the CatWISE2020 quasar sample deviate significantly from those obtained by a direct application of equation \ref{eq:kinematic_dipole} and lead to a larger deviation from the velocity predicted by the CMB dipole.

% \section*{Acknowledgements}
% The Acknowledgements section is not numbered. Here you can thank helpful
% colleagues, acknowledge funding agencies, telescopes and facilities used etc.
% Try to keep it short.

%%%%%%%%%%%%%%%%%%%%%%%%%%%%%%%%%%%%%%%%%%%%%%%%%%
\section*{Data Availability}
Data used in this article are already available in the public domain \citep{Secrest_2021} \footnote{\url{https://doi.org/10.5281/zenodo.4431089}}. 

%%%%%%%%%%%%%%%%%%%% REFERENCES %%%%%%%%%%%%%%%%%%

% The best way to enter references is to use BibTeX:

\bibliographystyle{mnras}
\bibliography{example} % if your bibtex file is called example.bib
%%%%%%%%%%%%%%%%%%%%%%%%%%%%%%%%%%%%%%%%%%%%%%%%%%

%%%%%%%%%%%%%%%%% APPENDICES %%%%%%%%%%%%%%%%%%%%%

% \appendix

% \section{Some extra material}
% If you want to present additional material which would interrupt the flow of the main paper,
% it can be placed in an Appendix which appears after the list of references.

%%%%%%%%%%%%%%%%%%%%%%%%%%%%%%%%%%%%%%%%%%%%%%%%%%

% Don't change these lines
\bsp	% typesetting comment
\label{lastpage}
\end{document}